\documentclass[11pt,a4paper]{article}
\usepackage[T1]{fontenc}

\usepackage[a4paper,top=2cm,bottom=2cm,left=3cm,right=3cm,marginparwidth=1.75cm]{geometry}
 
\usepackage{makecell}
\usepackage{amsmath}
\usepackage{graphicx}
\usepackage[hypertexnames=false]{hyperref}
\usepackage{longtable}   
\usepackage{booktabs}    
\usepackage{array}      
\usepackage{ragged2e}    
\usepackage{amssymb}
\usepackage{listings}
\usepackage{xcolor}
\lstdefinestyle{utilityspec}{
    language=Python,
    basicstyle=\footnotesize\ttfamily,
    backgroundcolor=\color{gray!8},
    keywordstyle=\color{green!40!black}\bfseries,
    commentstyle=\color{teal!70!black},
    stringstyle=\color{red!60!black},
    numbers=none,
    showstringspaces=false,
    keepspaces=true,
    columns=fullflexible,
    breaklines=true,
    breakatwhitespace=true,
    xleftmargin=0pt
}
\usepackage{pifont}
\newcommand{\cmark}{\ding{51}}
\usepackage{tabularx}
\usepackage{adjustbox}
\usepackage[most]{tcolorbox}
\usepackage{rotating}

\usepackage[affil-it]{authblk}

\makeatletter
\renewcommand{\section}{\@startsection {section}{1}{\z@}%
              {24pt}{12pt} {\large\scshape\bfseries}}

\renewcommand{\subsection}{\@startsection {subsection}{2}{\z@}%
             {12pt}{12pt}  {\itshape\bfseries}}
\makeatother

\usepackage{apacite}
\usepackage{natbib}

\defcitealias{van2019artificial}{Van Cranenburgh et al. (2019)}
\defcitealias{nova2025delphos}{Nova et al. (2025)}
\defcitealias{nova2025understanding}{Nova et al., 2025}

\title{\normalsize \textbf{Multitask Reinforcement Learning for Assisting Choice Model Specification}}
\author[1]{Gabriel Nova*}
\author[1,2]{Stephane Hess}
\author[1]{Sander van Cranenburgh}

\affil[1]{CityAI Lab, Transport and Logistics Group, Delft University of Technology, The Netherlands}
\affil[2]{Institute for Transport Studies and Choice Modelling Centre, University of Leeds, UK}

\date{\vspace{-5ex}}

\begin{document}
\maketitle

Discrete choice model specification is a time-consuming task in which modellers often specify and estimate multiple models while balancing goodness-of-fit, parsimony, and behavioural plausibility. We present Delphos, a multitask reinforcement learning framework that learns transferable specification strategies across transport choice datasets. Delphos frames model specification as a sequential decision-making problem in which it applies a sequence of modelling actions and receives feedback from an estimation environment based on model performance and convergence. To transfer modelling decisions across datasets with different sets of variables, Delphos represents utility specifications as sets of modelling terms using a DeepSet-Q architecture, allowing a shared specification policy to learn across multiple datasets. Trained on nine transport choice datasets, Delphos consistently outperforms independently trained single-task agents, indicating that sharing modelling experience improves learning efficiency and helps identify promising sequences of modelling decisions with fewer unsuccessful estimation attempts. When applied without further training to the unseen Swissmetro and Decisions datasets, the same agent identifies competitive specifications in less than 20 minutes on a standard CPU. It achieves a higher log-likelihood per observation than the VNS metaheuristic on Swissmetro and performance comparable to a published MNL specification developed by expert modellers on Decisions. These findings show that accumulating and reusing modelling experience enables Delphos to function as an intelligent assistant for discrete choice model specification. It reduces manual trial-and-error while allowing modellers to retain control over model diagnosis, refinement, and final selection.\\

\textbf{Keywords}: Assisted choice model specification; Deep reinforcement learning; Artificial intelligence 

\clearpage
\section{Introduction}
Discrete choice models (DCMs) are econometric frameworks widely used to understand and analyse individual choice behaviour, forecast demand, and evaluate policies across a wide range of application domains \cite{buckell2019should, mariel2021environmental, hess2024handbook, de2025value}. By representing the decision-making process underlying observed choices, DCMs provide insights into individuals’ preferences and the trade-offs they make between alternatives. These models can then be used to derive behavioural measures such as marginal utilities, elasticities, willingness-to-pay, and values of travel time. To obtain such behavioural insights, modellers typically specify competing choice models by making various modelling decisions, such as selecting explanatory variables, defining functional forms, and representing preference heterogeneity that reflect different behavioural assumptions \citep{van2022choice}. In practice, they estimate and evaluate specifications by balancing goodness-of-fit, parsimony, and behavioural plausibility, and then refine them through successive iterations \citepalias{nova2025understanding}. Consequently, specifying discrete choice models remains an iterative, cognitively demanding, and time-consuming process.\\

This trial-and-error workflow has motivated the development of assisted specification methods, which seek to support, speed up, or partially automate the specification process. Two broad classes of approaches can be distinguished. The first comprises metaheuristic methods that formulate model specification as a combinatorial search problem and use optimisation techniques to explore the space of possible models \citep{paez2022discrete, rodrigues2020bayesian, ortelli2021assisted}. These approaches iteratively construct candidate specifications from previously explored models and evaluate them using objective functions based on model fit and parsimony. Promising specifications are retained and modified through predefined search operators to generate new candidates. More recent approaches have incorporated behavioural constraints and grammar-based representations to improve the behavioural plausibility of the resulting specifications \citep{beeramoole2023extensive, haj5195530grammar, ghorbani2025enhanced}. Although these methods can efficiently explore model spaces, they rely on predefined search operators and do not explicitly accumulate knowledge about modelling decisions across specification problems.\\

The second class comprises machine-learning-based approaches that generate model specifications by learning from previous modelling exercises. Instead of relying on predefined search operators, these methods learn patterns from previously estimated specifications and use them to propose candidates that are likely to satisfy behavioural expectations and provide a good fit. For instance, \citet{sfeir2025can} explore the use of pre-trained large language models to generate utility specifications by drawing on knowledge and reasoning capabilities acquired from large text corpora. Similarly, \citetalias{nova2025delphos} formulate model specification as a sequential decision-making problem in which a reinforcement learning agent learns specification strategies through repeated interaction with an estimation environment. These approaches shift the focus from predefined search operators to learning-based systems that can use accumulated knowledge to assist the specification process.\\

However, existing assisted model specification methods struggle to reuse modelling experience across datasets. Human modellers, by contrast, rarely approach each specification problem independently. They draw on experience from previous modelling exercises and adapt it to new contexts. This is possible because many modelling decisions are informed by behavioural principles that generalise across datasets and application domains. For example, travel cost is expected to have a negative marginal utility, income interactions are commonly used to capture heterogeneity in cost sensitivity, and total, in-vehicle, and out-of-vehicle travel time can be treated as alternative representations of the same behavioural construct. Despite the transferability of this knowledge, metaheuristic methods typically discard information about previously explored specifications once the search terminates, offering little scope for cumulative learning. Machine-learning-based approaches retain knowledge from previous modelling tasks, but usually encode it in dataset-specific representations. Consequently, the specification process must effectively be repeated for each new dataset, even when relevant modelling knowledge could be transferred from related contexts.\\

To leverage modelling experience across datasets, we propose a multitask reinforcement learning framework for assisting choice model specification. By extending Delphos \citepalias{nova2025delphos} to a multitask setting, we enable a single agent to learn a transferable specification policy from multiple choice datasets. To support transfer across datasets with different attributes and socio-demographic variables, Delphos represents modelling decisions independently of the dataset-specific variables and contexts in which they are applied. Specifically, we extend Delphos with a DeepSet-Q network \citep{hugle2020dynamic} that encodes utility specifications as sets of modelling terms and learns a shared latent representation. Conditioned on the current modelling context, this representation enables the agent to select decisions that generalise across related datasets. Through training across datasets, the agent learns which modelling decisions tend to produce high-performing specifications in similar contexts and reuses this knowledge when specifying models for new datasets.\\

The main aim of the framework is to support modellers by using experience acquired across datasets to provide automated, data-driven suggestions for utility specifications. By directing the search towards high-performing and behaviourally plausible candidates, Delphos can reduce the cognitive and computational burden of manual trial-and-error and reduce the risk of misspecification. In practice, the modeller defines a catalogue of attributes, socio-demographic characteristics, non-linear transformations, and taste structures. Delphos then proposes and estimates candidate specifications and presents the successfully estimated models on the Pareto front. The modeller can compare, refine, and validate these candidates while retaining control over their behavioural interpretation and final selection. Overall, Delphos functions as an intelligent assistant that automates part of the specification workflow. It supports the development of utility functions from which willingness-to-pay measures, such as values of travel time savings used in cost--benefit analysis, can be derived.\\

The remainder of the paper is organised as follows. Section~\ref{sec:background} introduces the reinforcement learning concepts used in the paper. Section~\ref{sec:methodological_framework} presents our framework for assisted choice model specification across multiple datasets. Section~\ref{subsec:problem_formulation} formulates the specification task as a Markov decision process, Section~\ref{subsec:sharing_modelling_decisions_across_datasets} describes the multitask extension for sharing modelling decisions across datasets, and Section~\ref{subsec:training_eval_protocol} presents the training and evaluation protocol. Section~\ref{sec:experiment_setup} describes the experimental design and evaluation studies. Finally, Section~\ref{sec:results} presents the main results, and Section~\ref{sec:conclusions} concludes the paper.

\section{Background}\label{sec:background}
\subsection{Markov decision process} 
A Markov decision process (MDP) represents a sequential decision-making problem in which an agent interacts with an environment over time. The agent repeatedly observes the current state, selects an action, receives feedback from the environment, and transitions to a new state. Formally, an MDP is defined as the tuple $(\mathbf{S}, \mathbf{A}, \mathbf{P}, \mathbf{R}, \gamma)$, where $\mathbf{S}$ represents the possible states of the environment and $\mathbf{A}$ represents the actions available to the agent. The transition function $\mathbf{P}(s' \mid s,a)$ defines the probability of moving from state $s$ to a subsequent state $s'$ after taking action $a$. The reward function $\mathbf{R}(s,a)$ evaluates the outcome of taking action $a$ in state $s$. Finally, the discount factor $\gamma$ determines the relative importance of future and immediate rewards.

\begin{equation}\label{eq: value function}
V_{\pi}(s) = \mathbb{E}_{\pi} \left[ \sum_{t=0}^{\infty} \gamma^t r_{t+1}\;\middle|\; s_0 = s \right]
\end{equation}

The agent aims to maximise the cumulative reward obtained through its interactions with the environment. It follows a policy $\pi(a|s;\theta)$ that defines the probability of selecting action $a$ when the environment is in state $s$. The state-value function represents the expected cumulative reward obtained when starting from state $s$ and following policy $\pi$, as shown in Eq.~(\ref{eq: value function}). The optimal value function, denoted by $V^*(s)$, represents the maximum return achievable from state $s$.

\subsection{Reinforcement learning}
While an MDP formulates a sequential decision-making problem, reinforcement learning (RL) addresses that problem by learning a policy that maximises expected long-term rewards \citep{sutton2018reinforcement}. Through repeated interaction with the environment, the agent balances exploration and exploitation. Exploration provides additional knowledge about the environment, whereas exploitation uses that knowledge to select actions with high expected returns. At each decision step, the agent stores a transition containing the current state, selected action, received reward, and next state. These transitions are then used to refine the policy towards actions that are expected to yield higher returns.\\

Q-learning is commonly used to learn such a policy \citep{watkins1992q}. The agent estimates an action-value function, $Q(s,a)$, which represents the expected cumulative reward after taking action $a$ in state $s$, as shown in Eq.~(\ref{eq: action value}). These action values define a policy that selects the feasible action with the highest expected return in each state. Formally, the optimal policy maximises the action-value function over the set of feasible actions $\mathcal{A}(s)$, as shown in Eq.~(\ref{eq: greedy_policy}).

\begin{equation}\label{eq: action value}
    Q(s,a) = \mathbb{E} \left[r + \gamma \max_{a'} Q(s',a') \mid s,a \right]
\end{equation}

\begin{equation}
    \label{eq: greedy_policy}
    \pi^{*}(s) = \arg\max_{a \in \mathcal{A}(s)} Q^{*}(s,a)
\end{equation}

In practice, the action-value function cannot be represented explicitly when the state space is large. Deep Q-networks (DQNs; \citealp{mnih2015human}) address this limitation by approximating $Q(s,a;\theta)$ with a neural network parameterised by $\theta$. Rather than storing a value for every state--action pair, the network predicts action values directly from the state representation. To improve training stability, a DQN also maintains a target network $Q(s,a;\theta^-)$, whose parameters are periodically updated from the policy network and used to compute stable learning targets. This approach enables RL agents to operate in complex decision spaces where traditional tabular methods are infeasible \citep{plaat2022deep}.

\section{Methodological framework} \label{sec:methodological_framework}
This section introduces the proposed multitask reinforcement learning framework for assisting the choice model specification process across multiple datasets. We first formulate the specification process as a Markov decision process. We then extend the formulation to multiple datasets through a shared representation of utility specifications that enables transferable modelling decisions. Finally, we present the training and evaluation protocol used to assess learning and transferability.

\subsection{Problem formulation}\label{subsec:problem_formulation}
\begin{figure}[ht]
    \centering
    \includegraphics[width=\linewidth]{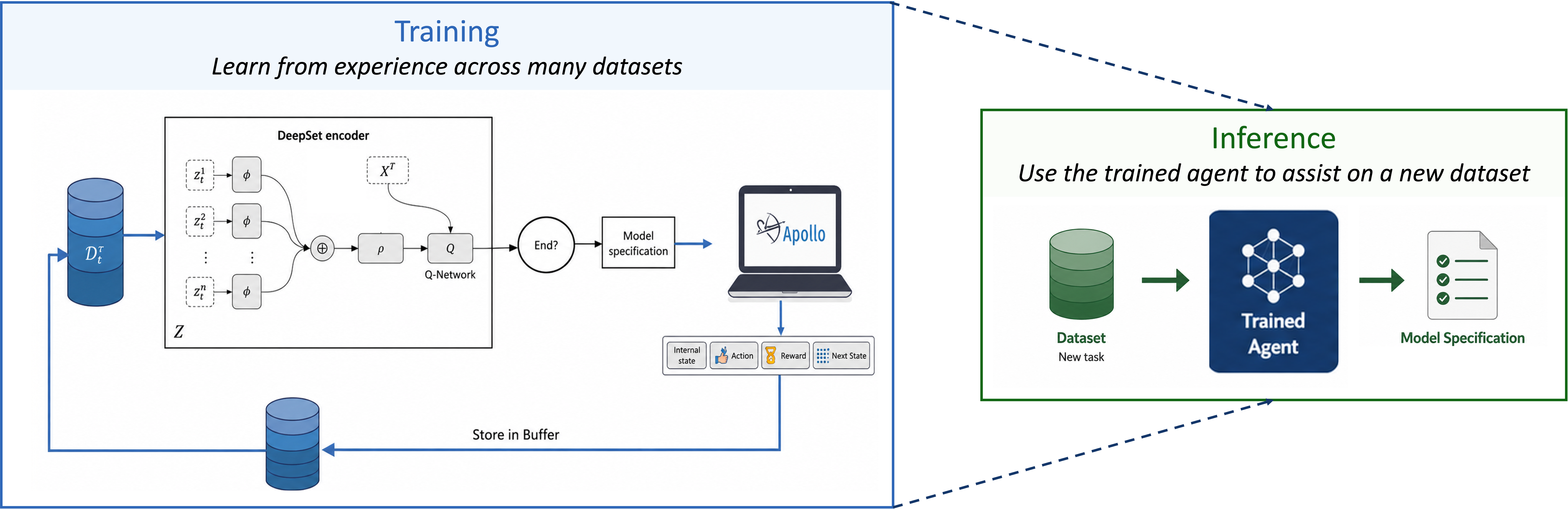}
    \caption{Delphos: a multitask reinforcement learning framework for assisting choice model specification across datasets. The agent learns to take a sequence of modelling actions to propose utility specifications, while the environment returns feedback based on model fit, complexity, and behavioural plausibility.}
    \label{fig: DQN_DCM_framework}
\end{figure}

We formulate discrete choice model specification as a Markov decision process in which an agent learns to build utility specifications through interaction with an estimation environment (Figure~\ref{fig: DQN_DCM_framework}). Delphos is the reinforcement learning agent that learns a policy for specifying high-performing choice models \citepalias{nova2025delphos}. At the start of each episode, Delphos uses a linear additive specification in which the available attributes enter linearly with generic coefficients and no covariate interactions. It then applies a sequence of modelling actions to modify these terms and propose a final candidate. The episode ends when the agent selects the \texttt{terminate} action. The environment then estimates the candidate and returns modelling outcomes used to compute a reward based on goodness-of-fit, convergence, and behavioural expectations.\\

The previous formulation considers a single specification problem, whereas our objective is to learn modelling strategies that transfer across related datasets. We therefore treat each choice dataset as a separate multinomial logit utility specification task $\tau$ sampled from a domain, such that $\tau \sim P(\tau)$ \citep{mcfadden1977modelling}. The domain is represented by a catalogue of modelling components that can be shared across tasks, including attributes, transformations, taste structures, and covariates. Each dataset $\tau$ uses a task-specific subset that is compatible with its available variables and feasible modelling terms. Consequently, each dataset induces its own MDP, with a task-specific state space $\mathcal{S}_{\tau}$, action space $\mathcal{A}_{\tau}$, transition function $P(s_{\tau,e+1}\mid s_{\tau,e},a_{\tau,e})$, and reward function $\mathcal{R}_{\tau}$.\\

To describe modelling decisions across specification problems, we define a \textbf{domain catalogue} containing all modelling components that may appear across the considered datasets. Formally, the domain catalogue is defined as
\begin{equation*}
    \mathcal{C} = \{\mathcal{K}, \mathcal{T}, \mathcal{G}, \mathcal{V}\},
\end{equation*}

where $\mathcal{K}$ is the set of available attributes that may enter the utility functions, such as travel time and travel cost. $\mathcal{T}$ is the set of transformations that can be applied to these attributes, including linear, logarithmic, and Box--Cox transformations. $\mathcal{G}$ is the set of taste structures, which determine whether a parameter is generic across alternatives or alternative-specific. $\mathcal{V}$ is the set of covariates used to capture observed heterogeneity. Each dataset uses a subset $\mathcal{C}_\tau$ and induces an MDP with a dataset-specific state space $\mathcal{S}_{\tau}$, action space $\mathcal{A}_\tau$, transition function $\mathcal{P}_\tau$, and reward function $\mathcal{R}_\tau$.\\

i) \textbf{State space ($\mathcal{S}_{\tau}$).} The state space comprises all feasible utility specifications that can be constructed from the dataset-specific catalogue $\mathcal{C}_{\tau}$. Each specification is represented as a set of modelling terms. Each term encodes a decision about how a variable enters the utility function. Specifically, a modelling term combines an attribute $k_l\in\mathcal{K}_{\tau}$ with its transformation $t_l\in\mathcal{T}_{\tau}$, taste structure $g_l\in\mathcal{G}_{\tau}$, and covariate interaction $v_l\in\mathcal{V}_{\tau}$. Formally, the state at interaction $e$ is

\begin{equation}\label{eq:state}
    s_e^\tau = \{x_l\}_{l=1}^{L_e}, \quad x_l = (k_l,t_l,g_l,v_l),
\end{equation}

where $L_e$ denotes the number of modelling terms in the specification. Attributes that are not included in the current specification are encoded as $t_l=g_l=v_l=0$.\\

To illustrate this representation, consider the following utility specification for the bus alternative. It contains four modelling terms: an alternative-specific constant; travel time with a logarithmic transformation and an alternative-specific coefficient; travel cost entering linearly with a generic coefficient that interacts with the two levels of the covariate $\text{cov1}$; and headway, which is excluded from the specification.

\begin{lstlisting}[style=utilityspec,frame=single]
V[["bus"]]=  asc_bus                              #(1, linear, specific,  none)
            + b_tt_bus    * log(TT_bus)           #(2, log,    specific, none)
            + b_tc_female * Female     * TC_bus   #(3, linear, generic,  female)
            + b_tc_male   * (1-Female) * TC_bus   #(3, linear, generic,  male)
            
\end{lstlisting}

Using the catalogue indices (e.g., attributes: 0=None, 1=ASC, 2=TT, 3=TC, 4=HE; transformations: 0=None, 1=Linear, 2=Log, 3=Box--Cox; taste structures: 0=None, 1=Generic, 2=Alternative-specific; covariates: 0=None, 1=COV1), the corresponding state is
\begin{equation}
    s_e^\tau= \{(1,1,2,0),\, (2,2,2,0),\, (3,1,1,1),\, (4,0,0,0)\}
\end{equation}

This representation preserves the behavioural decisions that define a utility specification while separating them from dataset-specific variable labels and availability. Consequently, variables that capture the same behavioural construct can be encoded consistently across datasets. Examples include total and in-vehicle travel time, or fare and travel cost. This shared representation provides the basis for transferring modelling experience from one specification problem to related tasks.\\

ii) \textbf{Action space ($\mathcal{A}_{\tau}$).} The action space defines the feasible operations that the agent can use to modify the current utility specification. Because each state is represented as a set of modelling terms, the agent operates directly on these terms. It can add a term, modify an existing term, or terminate the specification process. Formally, the action space is

\begin{equation}\label{eq:actions}
    a \in \{\text{add}(k,t,g,v),\ \text{change}(k,t,g,v),\ \text{terminate}\},
\end{equation}

where \textit{add} introduces a new modelling term, \textit{change} updates the transformation, taste structure, or covariate interaction of an existing term, and \textit{terminate} ends the specification process and triggers model estimation.\\

Continuing the example, suppose that the agent first selects \texttt{add}(HE, linear, generic, none), which introduces headway as a new modelling term. The updated specification is

\begin{tcolorbox}[
    colback=gray!8,
    colframe=black,
    boxrule=0.4pt,
    arc=0mm,
    left=3mm,
    right=3mm,
    top=2mm,
    bottom=2mm
]

\ttfamily
$s_e^0$ $\rightarrow$
\texttt{add}(HE, linear, generic, none)
$\rightarrow s_e^1$
\vspace{0.3em}
\begin{lstlisting}[style=utilityspec,frame=none]
V[["bus"]]=  asc_bus                            #(1, linear, specific,  none)
            + b_tt_bus    * log(TT_bus)         #(2, log,    specific, none)
            + b_tc_female * Female     * TC_bus #(3, linear, generic,  female)
            + b_tc_male   * (1-Female) * TC_bus #(3, linear, generic,  male)
            + b_he        * HE_bus              #(4, linear, generic,  none)
\end{lstlisting}
\end{tcolorbox}

The agent may then choose \texttt{change}(TT, Box-Cox, specific, none). This action replaces the logarithmic transformation of travel time with a Box--Cox transformation while preserving the remaining modelling decisions. The resulting specification is
\begin{tcolorbox}[
    colback=gray!8,
    colframe=black,
    boxrule=0.4pt,
    arc=0mm,
    left=3mm,
    right=3mm,
    top=2mm,
    bottom=2mm
]

\ttfamily
$s_e^1$ $\rightarrow$
\texttt{change}(TT, box-cox, specific, none)
$\rightarrow s_e^2$
\vspace{0.3em}
\begin{lstlisting}[style=utilityspec,frame=none]
V[["bus"]]=  asc_bus                            #(1, linear, specific,  none)
            + b_tt_bus    * BoxCox(TT_bus)      #(2, box-cox, specific, none)
            + b_tc_female * Female     * TC_bus #(3, linear, generic,  female)
            + b_tc_male   * (1-Female) * TC_bus #(3, linear, generic,  male)
            + b_he        * HE_bus              #(4, linear, generic,  none)
\end{lstlisting}
\end{tcolorbox}

This formulation decomposes utility specification into a sequence of modelling decisions, whereby the agent progressively constructs a utility specification by adding and modifying modelling terms until it selects the \texttt{terminate} action.  The previous example thus corresponds to a trajectory of state transitions, in which the agent progressively updates the specification through successive modelling actions before submitting the final candidate for estimation.

\begin{tcolorbox}[
colback=gray!8,
colframe=black,
boxrule=0.4pt,
arc=0mm,
left=3mm,
right=3mm,
top=2mm,
bottom=2mm
]
\ttfamily
s$_0$\\[0.3em]
\hspace*{1em}$\hookrightarrow$ \texttt{add}(HE, linear, generic, none)
$\rightarrow$ s$_1$\\[0.3em]
\hspace*{3em}$\hookrightarrow$ \texttt{change}(TT, box--cox, specific, none)
$\rightarrow$ s$_2$\\[0.3em]
\hspace*{5em}$\hookrightarrow$ \texttt{terminate}
$\rightarrow$ s$_3$\\[0.3em]
\hspace*{7em}$\hookrightarrow$ environment estimates $s_3$
$\rightarrow$ reward
\end{tcolorbox}

To ensure that each decision contributes meaningfully to the specification process, the available actions are dynamically restricted. The agent cannot select modelling components that are unavailable in the dataset, actions that immediately reverse previous modifications, or actions already selected in the current episode. An action-masking mechanism implements these restrictions and ensures that the agent explores only feasible and non-redundant specification trajectories \citep{huang2020closer}.\\

iii) \textbf{Reward signal ($\mathcal{R}_{\tau}$).} The reward signal evaluates the quality of the utility specification at the end of an episode. Through delayed credit assignment, this feedback is propagated to the preceding sequence of modelling decisions. The agent can therefore learn which decisions improve model performance over repeated trials. In a multitask setting, however, goodness-of-fit measures such as log-likelihood cannot be used directly because their magnitude depends on the sample size, model complexity, and baseline fit of each dataset. Similar modelling decisions may therefore produce substantially different rewards across datasets. This could bias learning towards datasets with larger absolute improvements in log-likelihood. To provide a stable signal across heterogeneous datasets, we define the reward as the per-observation improvement in log-likelihood relative to a baseline model:
\begin{equation}\label{eq:reward}
    r = \tanh\!\Big(\frac{LL_{\hat{\beta}} - LL_{baseline}}{N_{obs}}\Big),
\end{equation}

Normalising by the number of observations makes the reward comparable across datasets of different sizes. The $\tanh$ transformation bounds the reward within $(-1,1)$ and improves numerical stability in a similar way to reward clipping \citep{mnih2015human,hessel2019multi}. We use the null model as the baseline because it provides a lower bound for theoretically valid maximum-likelihood specifications \citep{mokhtarian2016discrete}. Specifications that fail to converge receive a fixed penalty of $-1$. The reward can also accommodate richer modelling objectives through penalties or incentives based on behavioural expectations. For example, it could discourage specifications with implausible parameter signs, such as positive travel-time or travel-cost sensitivities, statistically insignificant parameters, or estimation issues (\citetalias{nova2025delphos}).\\

iv) \textbf{Environment.} The environment provides the interface between the reinforcement learning agent and the choice model estimation framework. After the agent selects the \texttt{terminate} action, the environment translates the final state into a complete utility specification and estimates it on the corresponding dataset using Apollo \citep{hess2019apollo}. The resulting outcomes, including goodness-of-fit measures, convergence diagnostics, and parameter estimates, are used to compute the reward in Eq.~(\ref{eq:reward}).

\subsection{Learning transferable modelling decisions}\label{subsec:sharing_modelling_decisions_across_datasets} 
To enable a single agent to learn and reuse modelling experience across choice datasets, we extend Delphos to a multitask reinforcement learning framework. Instead of training a separate agent for each specification problem, Delphos learns jointly across related problems. It learns both which modelling decisions tend to produce high-performing utility specifications and the contexts in which those decisions are appropriate. A single policy, however, requires a common representation of utility specifications despite differences in the number and composition of their modelling terms.\\

To address this heterogeneity, we implement Delphos as a DeepSet-Q network \citep{hugle2020dynamic}. DeepSets encode unordered sets with varying numbers of elements as fixed-length latent representations \citep{zaheer2017deep}. They are therefore well suited to utility specifications, whose modelling terms have no intrinsic order and may differ in number and composition. In the single-task framework of \citet{nova2025delphos}, each specification is represented as a fixed-length vector. By contrast, we represent utility specifications as sets of modelling terms. Specifications from different datasets can therefore be represented consistently and processed by a single policy.\\

\begin{figure}[h]
        \centering
        \includegraphics[width=0.5\linewidth]{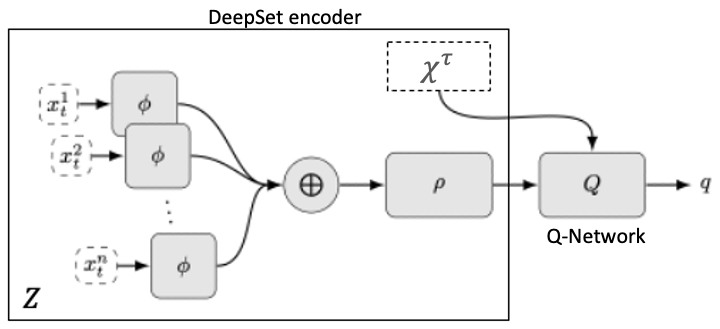}
        \caption{DeepSet-Q network (adapted from \cite{hugle2020dynamic}). It combines a specification encoder with a context-dependent Q-network. Each modelling term is embedded by a shared network $\phi(\cdot)$ and then aggregated through pooling $\rho(\cdot)$ to obtain a latent representation of the specification.}
        \label{fig:DeepSet-Q}
\end{figure} 

Figure~\ref{fig:DeepSet-Q} shows the proposed architecture, which combines a specification encoder with a context-dependent Q-network. The architecture separates the representation of a utility specification from the modelling context in which it is applied. The first component is the DeepSet encoder \citep{zaheer2017deep}, which maps the set of modelling terms $s_e^\tau=\{x_l\}_{l=1}^{L_e}$ to a fixed-length latent representation. Each term $x_l$ is embedded using a shared neural network $\phi(\cdot)$ and then aggregated through a permutation-invariant pooling operation,

\begin{equation}
    Z(s_e^\tau)=\rho\!\left(\sum_{l=1}^{L_e}\phi(x_l)\right),    
\end{equation}

where $Z(s_e^\tau)$ is the latent representation of the utility specification. Because the aggregation is invariant to the order and number of modelling terms, specifications from different datasets are mapped to a common latent space.\\

The specification embedding alone does not identify which modelling terms are available for the current dataset. Delphos therefore uses a task-specific context vector $x^\tau$, which encodes the availability of attributes, transformations, taste structures, and covariates in the dataset-specific catalogue $\mathcal{C}_\tau$. The specification embedding $Z(s_e^\tau)$ and context vector $x^\tau$ are concatenated and provided as input to a deep Q-network \citep{mnih2015human},

\begin{equation}
    q= Q\!\left(Z(s_e^\tau) \,||\, x^\tau\right),
\end{equation}

which estimates the value of each feasible modelling decision in the current state. By combining a shared specification representation with a task-specific context, Delphos learns a single policy that can be applied across related datasets.

\subsection{Multitask training and evaluation}\label{subsec:training_eval_protocol} 
Delphos is trained jointly across all specification tasks using a single DeepSet-Q network whose parameters are shared across datasets. During each episode, a dataset $\tau$ is sampled at random, and the agent selects modelling actions until it reaches the \texttt{terminate} action. The environment estimates the resulting utility specification and assigns a reward according to Eq.~(\ref{eq:reward}). Each interaction generates a transition $(s_e^\tau,a_e^\tau,r_e^\tau,s_{e+1}^\tau)$, which is stored in a shared experience replay buffer. The buffer therefore contains modelling experience from all training datasets and is used to update the network.\\

The network parameters are updated following the deep Q-learning framework \citep{mnih2015human}. At each optimisation step, a mini-batch $\mathcal{B}$ of transitions is sampled from the buffer, and the temporal-difference loss is minimised,

\begin{equation}
    \mathcal{L}(\theta)= \frac{1}{|\mathcal{B}|}\sum_{b=1}^{|\mathcal{B}|} \left[Q_{\mathrm{target}}(s',a';\theta^-) - Q(s,a;\theta)\right]^2,
\end{equation}

where $\theta$ and $\theta^-$ denote the parameters of the policy and target networks, respectively. The loss compares the estimated value of a modelling decision with a target that combines the observed reward and expected value of subsequent decisions. By minimising this error, Delphos learns to favour decisions that improve the current utility specification and lead to more promising sequences of decisions over time.\\

To prevent datasets with longer specification trajectories from dominating optimisation, we sample balanced mini-batches from the replay buffer so that each task contributes similarly to the parameter updates \citep{sutton2018reinforcement}. During training, $\epsilon$-greedy exploration is gradually reduced, shifting the agent from exploration towards decisions associated with high-performing specifications. This design follows the purpose of experience replay in deep reinforcement learning \citep{schaul2015prioritized} and provides balanced \citep{sodhani2021multi} and diverse \citep{ross2010efficient} exposure across tasks.\\

During inference, the trained agent is applied directly to unseen datasets without further training. Given the task-specific catalogue $\mathcal{C}_\tau$, Delphos selects a sequence of modelling decisions before submitting a utility specification for estimation. By repeating this process, Delphos proposes and estimates a set of high-performing candidates. Modellers can then inspect and compare the resulting models using criteria such as goodness-of-fit, behavioural plausibility, and parameter significance. Delphos thus acts as a specification assistant that reuses experience from related problems while adapting its decisions to previously unseen datasets.

\section{Experiments} \label{sec:experiment_setup}
This section evaluates the proposed multitask reinforcement learning framework. We first introduce the choice datasets used for training and inference (Subsection~\ref{subsection: datasets}). We then describe the metrics used to assess learning and transferability (Subsection~\ref{subsection: evaluation metrics}). Finally, we present the studies used to evaluate whether modelling experience transfers across specification problems and whether the specifications generated for unseen datasets are behaviourally plausible (Subsection~\ref{subsection: evaluation studies}).

\subsection{Datasets}\label{subsection: datasets}
Table~\ref{tab:choice-datasets} summarises the transport choice datasets used in this paper. The first nine are used for training, while Decisions and Swissmetro are held out for inference. Each dataset defines a different specification problem, with its own choice context, alternatives, available attributes, socio-demographic variables, and sample size. Although the datasets differ in their observed variables and experimental settings, they share behavioural concepts such as travel time and travel cost, among others.

\newtcolorbox{tablebox}{
  enhanced,
  colback=gray!8,
  colframe=black,
  boxrule=0.4pt,
  arc=0mm,
  left=1.5mm,
  right=1.5mm,
  top=1.5mm,
  bottom=1.5mm,
  boxsep=0mm,
  width=\textwidth
}
\begin{table*}[t]
\centering
\caption{Choice modelling tasks used for multitask training and inference.}
\label{tab:choice-datasets}
\begin{tablebox}
\centering
\scriptsize
\setlength{\tabcolsep}{2.2pt}
\renewcommand{\arraystretch}{1.08}
\begin{adjustbox}{max totalsize={\linewidth}{0.8\textheight},keepaspectratio,center}
\begin{tabular}{@{}l*{11}{c}@{}}
\textbf{Task $\tau$}
& $\tau_1$
& $\tau_2$
& $\tau_3$
& $\tau_4$
& $\tau_5$
& $\tau_6$
& $\tau_7$
& $\tau_8$
& $\tau_9$
& $\tau_{10}$
& $\tau_{11}$\\
\midrule
\textbf{Training datasets}
& \cmark & \cmark & \cmark & \cmark & \cmark & \cmark & \cmark & \cmark & \cmark &   &  \\

\midrule
\multicolumn{12}{@{}l}{\textbf{Choice attributes}}\\

Travel time
& \cmark & \cmark & \cmark & \cmark & \cmark &  & \cmark & \cmark & \cmark & \cmark & \cmark \\

Travel cost
& \cmark & \cmark & \cmark & \cmark & \cmark & \cmark & \cmark & \cmark & \cmark & \cmark & \cmark \\

Out-of-vehicle time
& \cmark &  &  &  & \cmark & \cmark & \cmark & \cmark &  & \cmark & \cmark \\

Transfers
&  & \cmark & \cmark &  & \cmark &  & \cmark & \cmark &  &  &  \\

Service quality
& \cmark &  & \cmark &  &  &  &  &  &  &  & \cmark \\

Reliability
&  &  &  &  & \cmark &  & \cmark &  &  &  &  \\

\midrule
\multicolumn{12}{@{}l}{\textbf{Socio-demographic variables}}\\

Gender
& \cmark &  & \cmark & \cmark &  & \cmark & \cmark & \cmark & \cmark & \cmark & \cmark \\

Income
& \cmark & \cmark &  & \cmark &  & \cmark &  & \cmark & \cmark & \cmark & \cmark \\

Age
&  &  & \cmark & \cmark &  & \cmark & \cmark & \cmark & \cmark & \cmark & \cmark \\

Purpose
&  & \cmark & \cmark & \cmark & \cmark &  & \cmark &  &  & \cmark & \cmark \\

Car access
&  & \cmark &  &  &  &  & \cmark & \cmark &  & \cmark &  \\

Business
& \cmark & \cmark &  &  &  &  &  &  &  &  & \cmark \\

Education
&  &  & \cmark &  &  &  &  & \cmark & \cmark & \cmark &  \\

\midrule
\multicolumn{12}{@{}l}{\textbf{Dataset characteristics}}\\

Panel data
& \cmark & \cmark & \cmark & \cmark & \cmark & \cmark &  & \cmark & \cmark &  & \cmark \\

Alternatives
& 4 & 2 & 2 & 2 & 3 & 3 & 4 & 2 & 3 & 6 & 3 \\

Sample size
& 7000 & 3492 & 1511 & 52488 & 6570 & 1576 & 81086 & 1521 & 850 & 9356 & 5409 \\

$LL_0$
& -8196 & -2420 & -1047 & -36381 & -5885 & -1731 & -112409 & -1054 & -966 & -11233 & -5548 \\

$LL_{\mathrm{linear}}$
& -5761 & -1665 & -732 & -33026 & -4734 & -1121 & -71099 & -812 & -924 & -4294 & -4346 \\

\midrule
&
\rotatebox{90}{\cite{hess2019apollo}} &
\rotatebox{90}{\cite{axhausen2008income}} &
\rotatebox{90}{\cite{biogeme_Optima}} &
\rotatebox{90}{\cite{ramjerdi2010value}} &
\rotatebox{90}{\cite{arentze2013travelers}} &
\rotatebox{90}{\cite{ibeas2014modelling}} &
\rotatebox{90}{\cite{hillel2018recreating}} &
\rotatebox{90}{\cite{biogeme_Optima}} &
\rotatebox{90}{\citetalias{van2019artificial}} &
\rotatebox{90}{\cite{calastri2020we}} &
\rotatebox{90}{\cite{bierlaire2001acceptance}} \\
\end{tabular}
\end{adjustbox}
\end{tablebox}
\end{table*}

\subsection{Evaluation metrics}\label{subsection: evaluation metrics}

The experiments evaluate two aspects of the proposed framework: (i) its ability to learn sequences of modelling decisions that produce high-performing utility specifications during multitask training and (ii) its ability to transfer these strategies to unseen specification problems during inference.\\

During training, Delphos sequentially proposes utility specifications and receives feedback from the estimation environment. We evaluate learning performance using the average reward, maximum reward, and area under the learning curve. These metrics measure the average quality of the modelling trajectories, the quality of the best specification found, and overall learning efficiency, respectively. To characterise exploration, we also report the convergence rate, defined as the proportion of estimated specifications that converge successfully, and the proportion of novel specifications explored during training.\\

During inference, the trained agent is applied to unseen datasets without further training. We use the same metrics to evaluate the generated specifications and the transferability of the learnt policy. We also derive the Pareto front to examine the trade-off between model fit and complexity and to characterise the models that Delphos tends to propose.

\subsection{Evaluation studies}\label{subsection: evaluation studies}

To evaluate Delphos’ ability to learn transferable modelling strategies, we examine whether it can use experience across datasets to improve sample efficiency
and whether the generated specifications balance model fit, parsimony, and behavioural plausibility.\\

First, we evaluate whether the proposed multitask framework improves Delphos’ ability to learn and transfer modelling experience across specification problems. We compare the proposed DeepSet-Q architecture with the single-task DQN framework introduced by \citetalias{nova2025delphos}. All agents are trained for 10,000 episodes using the hyperparameter configuration reported in Appendix~\ref{tab:hyperparameters}. The multitask agent is trained jointly across all training datasets, whereas each single-task agent is trained independently on its corresponding dataset. We evaluate learning and transfer using the metrics introduced in Subsection~\ref{subsection: evaluation metrics}.\\

Second, we assess the quality of the utility specifications generated by Delphos for unseen datasets. We compare the Pareto-optimal specifications identified by Delphos with benchmark models from the original studies and with other assisted specification approaches. We consider model performance, parsimony, and behavioural plausibility to determine whether the transferred policy produces competitive models that are consistent with established choice modelling practice.

\section{Results}\label{sec:results}
This section presents the results of the two evaluation studies. We first examine learning during multitask training and then assess the utility specifications generated for unseen datasets.

\subsection{Learning transferable specification strategies}
\begin{figure}[ht]
  \centering
  \includegraphics[width=0.98\linewidth]{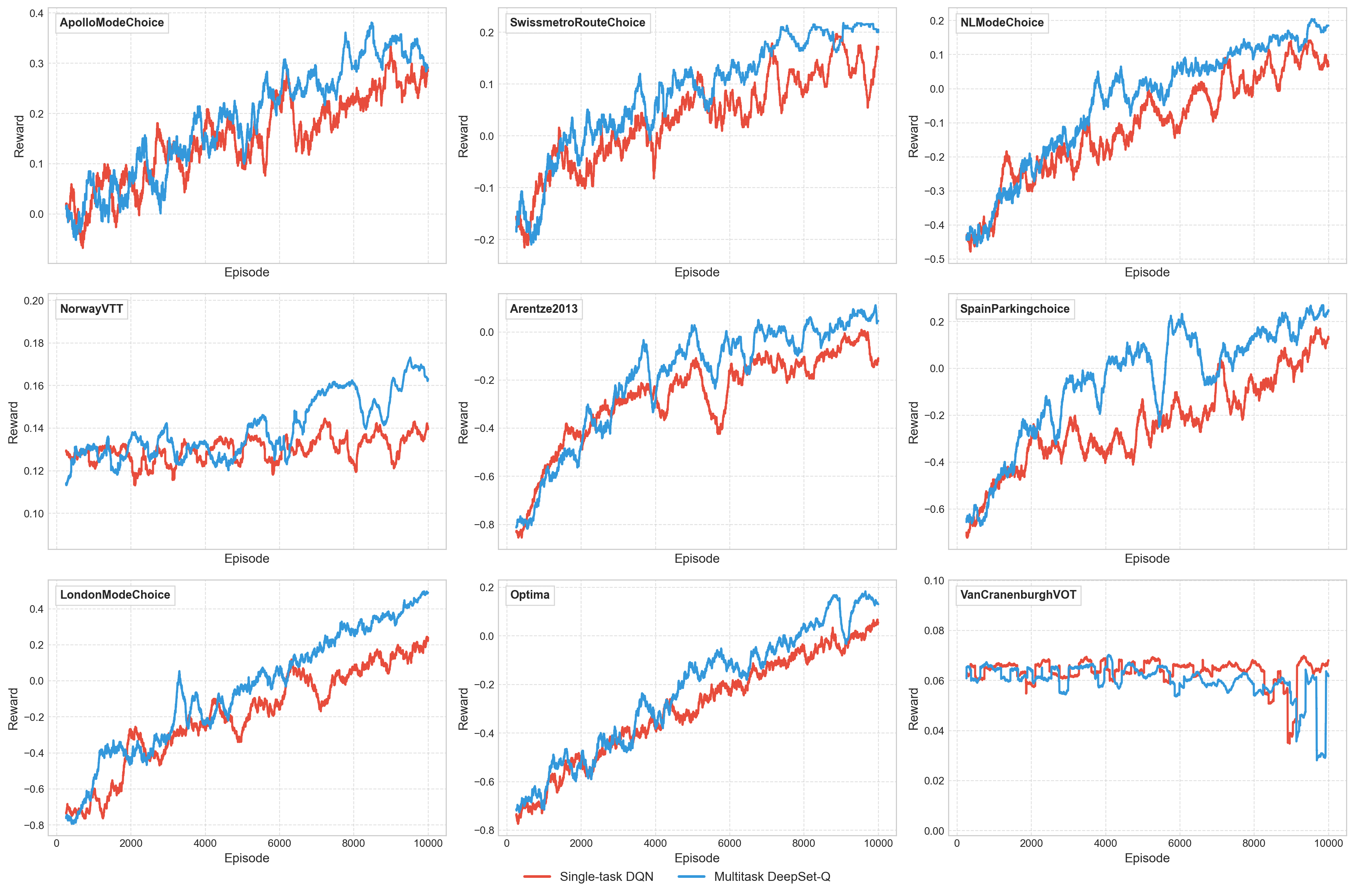}
  \caption{Learning curves of the single-task DQN agents and the proposed multitask DeepSet-Q agent across the training datasets.}
  \label{fig:learning_curve}
\end{figure}

Figure~\ref{fig:learning_curve} compares the learning curves of the proposed multitask DeepSet-Q agent with those of the single-task DQN agents across the training datasets. For most datasets, both approaches achieve progressively higher rewards during training. However, the multitask agent generally learns faster and reaches higher reward levels. This suggests that updating a single policy with experience from multiple specification problems improves its ability to generalise across datasets.\\

Learning performance varies across datasets. The gains are larger for LondonModeChoice, SpainParkingChoice, and ApolloModeChoice, and the differences are smaller for NorwayVTT and VanCranenburghVOT. One possible explanation is the complexity of the underlying specification problem. Datasets with more attributes and feasible covariate interactions define larger specification spaces and require longer sequences of modelling decisions. In these settings, experience from related datasets provides greater guidance during specification. By contrast, NorwayVTT and VanCranenburghVOT contain only two attributes: travel time and travel cost. Their smaller specification and action spaces make transferred modelling experience less beneficial.\\

Table~\ref{tab:representation_learning} summarises the performance of the single-task DQN agents and the proposed multitask DeepSet-Q agent across the training datasets. The multitask agent generally achieves higher mean rewards, maximum rewards, and areas under the learning curve. These improvements indicate that sharing modelling experience across specification problems enables Delphos to learn more effective policies. The main exception is VanCranenburghVOT, for which the two approaches perform similarly. This result is consistent with its smaller specification space.\\

\begin{table}[ht]
\centering
\caption{Learning performance on the training datasets.}
\label{tab:representation_learning}
\scriptsize
\begin{tabular}{llrrrrr}
\toprule
\textbf{Dataset} &
\textbf{Agent} &
\makecell{\textbf{Mean}\\\textbf{Reward}} &
\makecell{\textbf{Max}\\\textbf{Reward}} &
\textbf{AUC} &
\makecell{\textbf{Novel}\\\textbf{Specs (\%)}} &
\makecell{\textbf{Estimable}\\\textbf{Specs (\%)}} \\
\midrule

ApolloModeChoice
& DQN
& 0.151
& 0.334
& 1466
& \textbf{93.66}
& 82.74 \\
& \textbf{DeepSet-Q}
& \textbf{0.187}
& \textbf{0.434}
& \textbf{1839}
& 91.75
& \textbf{85.56} \\

\addlinespace

SwissmetroRouteChoice
& DQN
& 0.033
& 0.219
& 291
& \textbf{95.86}
& 84.62 \\
& \textbf{DeepSet-Q}
& \textbf{0.082}
& \textbf{0.220}
& \textbf{761}
& 92.87
& \textbf{88.52} \\

\addlinespace

NLModeChoice
& DQN
& -0.111
& \textbf{0.216}
& -1161
& \textbf{96.27}
& 72.80 \\
& \textbf{DeepSet-Q}
& \textbf{-0.042}
& 0.205
& \textbf{-497}
& 89.54
& \textbf{78.61} \\

\addlinespace

NorwayVTT
& DQN
& 0.130
& 0.168
& 1300
& \textbf{26.24}
& \textbf{99.84} \\
& \textbf{DeepSet-Q}
& \textbf{0.140}
& \textbf{0.173}
& \textbf{1394}
& 22.66
& \textbf{99.84} \\

\addlinespace

Arentze2013
& DQN
& -0.273
& 0.008
& -2825
& \textbf{77.90}
& 62.01 \\
& \textbf{DeepSet-Q}
& \textbf{-0.199}
& \textbf{0.111}
& \textbf{-2096}
& 63.32
& \textbf{68.35} \\

\addlinespace

SpainParkingChoice
& DQN
& -0.237
& 0.176
& -2473
& \textbf{17.99}
& 54.91 \\
& \textbf{DeepSet-Q}
& \textbf{-0.058}
& \textbf{0.270}
& \textbf{-698}
& 13.64
& \textbf{68.21} \\

\addlinespace

LondonModeChoice
& DQN
& -0.196
& \textbf{0.518}
& -2086
& \textbf{96.74}
& 52.62 \\
& \textbf{DeepSet-Q}
& \textbf{-0.056}
& 0.497
& \textbf{-716}
& 91.81
& \textbf{61.22} \\

\addlinespace

Optima
& DQN
& -0.297
& 0.181
& -3060
& \textbf{96.47}
& 57.61 \\
& \textbf{DeepSet-Q}
& \textbf{-0.228}
& \textbf{0.183}
& \textbf{-2376}
& 89.27
& \textbf{63.83} \\

\addlinespace

VanCranenburghVOT
& \textbf{DQN}
& \textbf{0.064}
& 0.086
& \textbf{635}
& \textbf{27.77}
& \textbf{99.73} \\
& DeepSet-Q
& 0.060
& \textbf{0.089}
& 597
& 25.20
& 99.72 \\

\bottomrule
\end{tabular}
\end{table}

From a reinforcement learning perspective, the higher rewards are accompanied by larger areas under the learning curve, which suggests that Delphos reaches high-performing specifications earlier during training. Notably, these improvements are obtained while exploring a slightly smaller proportion of novel specifications. This suggests that modelling experience acquired from related datasets makes the search more sample-efficient,  which allows the agent to identify promising sequences of modelling decisions with fewer unsuccessful specification trials.\\

This behaviour is also reflected from a choice modelling perspective. Although the multitask agent explores fewer unique specifications, it consistently produces a larger proportion of successfully estimated models across almost all datasets. This suggests that the transferred modelling experience not only improves model fit but also guides the search towards specifications that are more likely to be successful estimated.


\subsection{Application to unseen datasets}

\subsubsection{Swissmetro}
\begin{figure}[ht]
  \centering
  \includegraphics[width=0.85\linewidth]{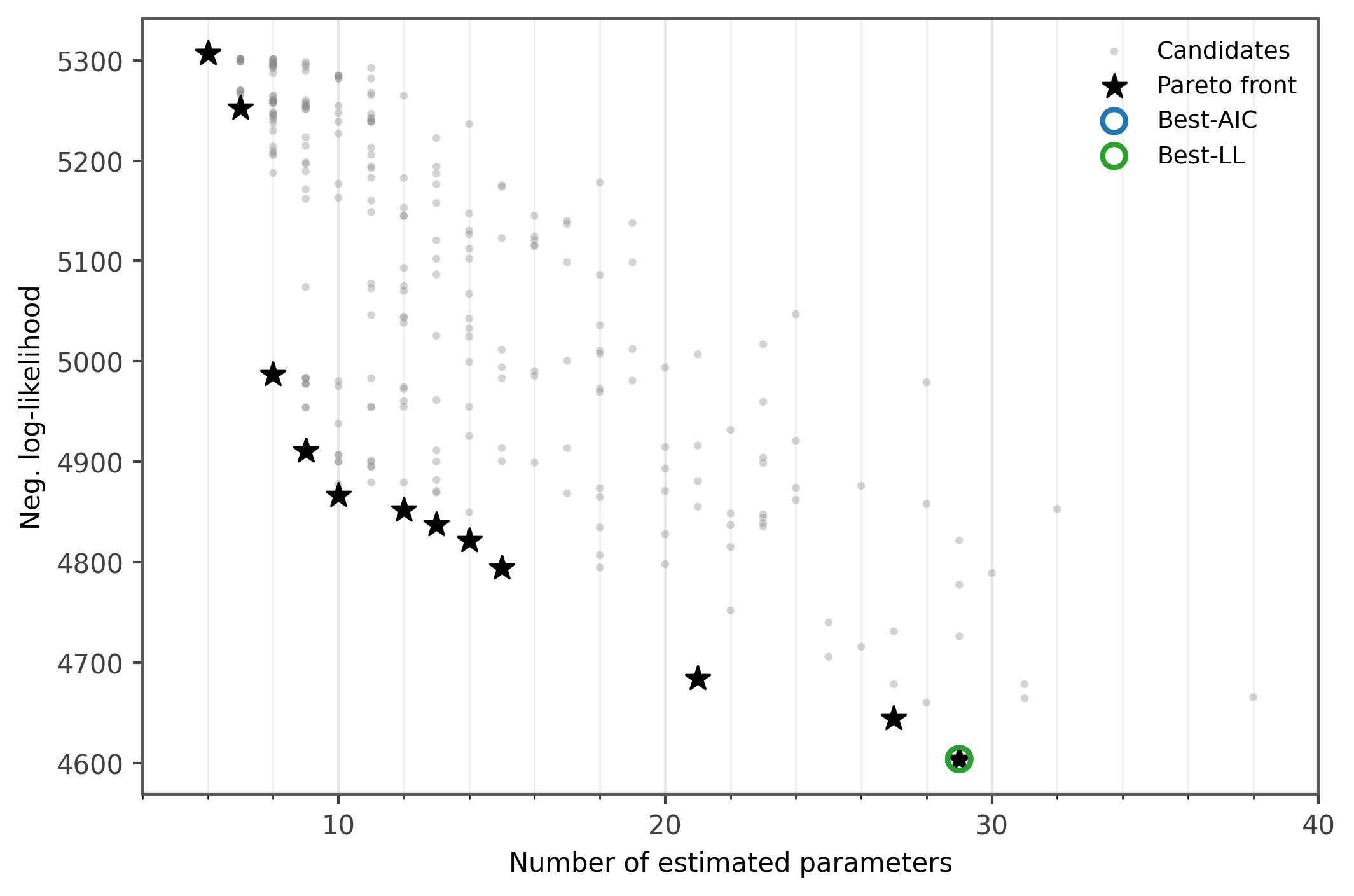}
  \caption{Pareto front of Delphos candidates on the Swissmetro dataset. Grey dots show all converged specifications proposed during inference. Stars indicate the non-dominated candidates, and circles show the specifications with the best AIC and log-likelihood values.}
  \label{fig:Pareto-Swissmetro}
\end{figure}

When applied to the unseen Swissmetro dataset ($\tau_{11}$; \citep{bierlaire2001acceptance}), Delphos proposes 283 candidate models within a computational budget of 20 minutes on a laptop CPU. Figure~\ref{fig:Pareto-Swissmetro} shows all converged candidates and the resulting Pareto front, whose non-dominated models balance model performance and parsimony. Thirteen of the converged specifications are non-dominated. Within the computational budget, Delphos identifies a best specification with a log-likelihood of $-0.681$ per observation. This result improves on both the linear baseline ($LL_{\mathrm{linear}}=-1.029$) and the value of $-0.72$ per observation reported for the VNS assisted specification method by \cite{ortelli2021assisted}.\\

\begin{table}[ht]
\centering
\caption{Best specification identified by Delphos on Swissmetro.}
\label{tab:swissmetro_spec_and_fit}
\small
\begin{tabular}{lcccc}
\hline
Modelling term & Included & Taste & Transformation & Interaction \\
\hline
ASC         & \checkmark & Specific & --       &  \\
Travel time & \checkmark & Specific & Log & Age \\
Travel cost & \checkmark & Specific & Log   & Gender \\
Headway     & \checkmark & Generic  & Linear      & Gender \\
Seat type   & \checkmark & Generic  & Linear      & Income \\
\hline
LL($0$) & \multicolumn{4}{r}{-6,964.66} \\
LL(final) & \multicolumn{4}{r}{-4,603.96} \\
AIC & \multicolumn{4}{r}{9,265.92} \\
BIC & \multicolumn{4}{r}{9,463.70} \\
Rho-squared & \multicolumn{4}{r}{0.34} \\
Adj.\ Rho-squared & \multicolumn{4}{r}{0.34} \\
Observations & \multicolumn{4}{r}{6,768} \\
Number of parameters & \multicolumn{4}{r}{29} \\\hline
$LL/N$ -- Delphos & \multicolumn{4}{r}{-0.68} \\
$LL/N$ -- VNS \cite{ortelli2021assisted} & \multicolumn{4}{r}{-0.72} \\
\hline
\end{tabular}
\end{table}

\clearpage
Table~\ref{tab:swissmetro_spec_and_fit} reports the specification identified by Delphos and its fit statistics. The model includes constants specific to each alternative and relevant attributes with either generic parameters or parameters that vary across alternatives. It also introduces non-linear transformations and meaningful covariate interactions. These results suggest that Delphos can transfer modelling decisions to an unseen dataset and identify a competitive, behaviourally plausible specification within a limited computational budget. Inference runs on a standard laptop CPU, making the approach practical for analysts seeking to reduce specification effort and for researchers studying reusable modelling strategies.

\subsubsection{Decisions}
\begin{figure}[ht]
  \centering
  \includegraphics[width=0.85\linewidth]{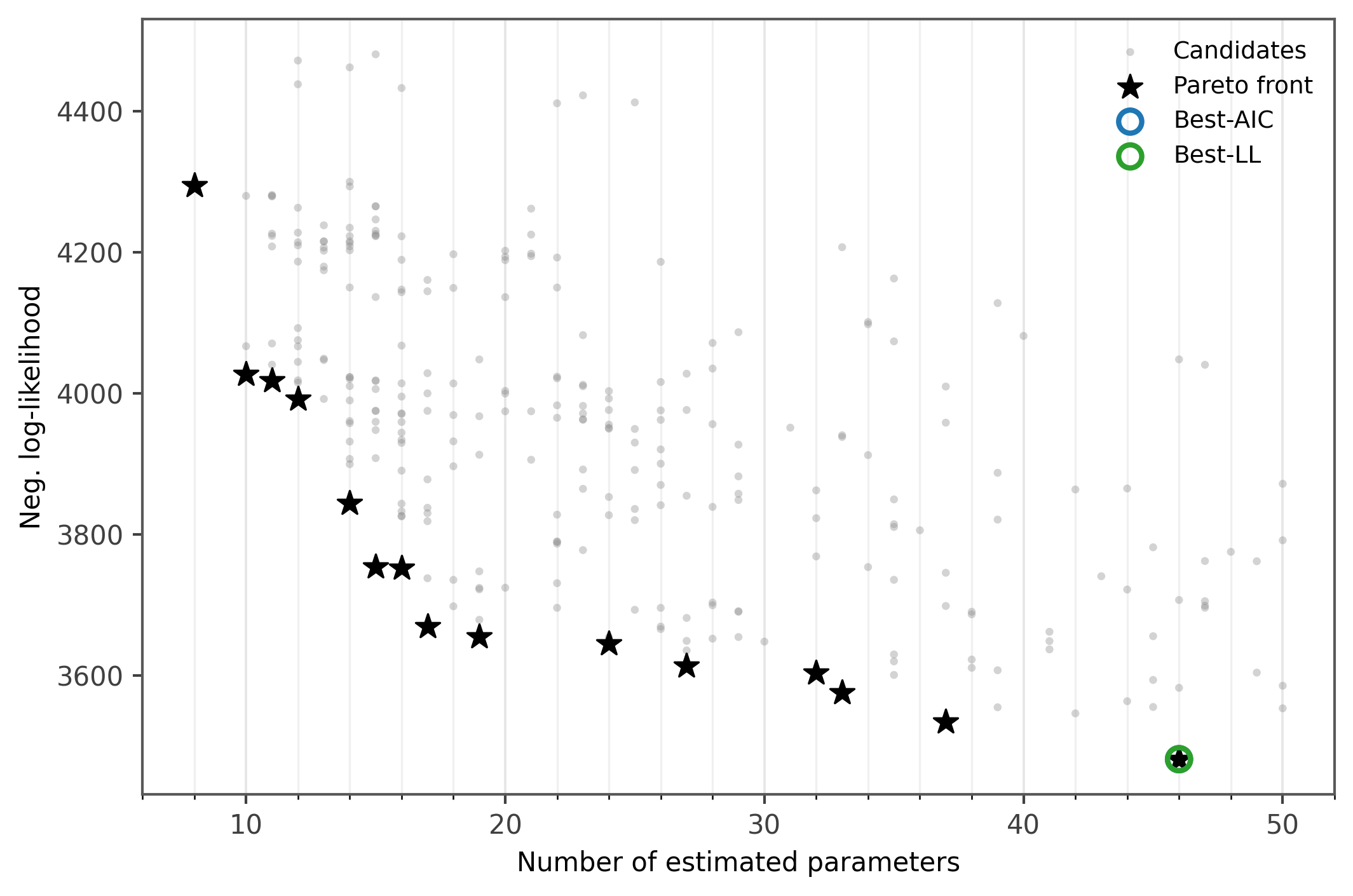}
  \caption{Pareto front of Delphos candidates on the Decisions dataset. Grey dots show all converged specifications proposed during inference. Stars indicate the non-dominated candidates, and circles show the specifications with the best AIC and log-likelihood values.}
  \label{fig:Pareto-Decisions}
\end{figure}

We apply the same trained multitask agent to the unseen Decisions dataset ($\tau_{10}$; \citep{calastri2020we}) using the same 20-minute computational budget. Figure~\ref{fig:Pareto-Decisions} shows all converged specifications and the resulting Pareto front. Without further training, Delphos identifies a set of non-dominated models. This result suggests that the learnt policy continues to select effective modelling actions for a different and unseen specification problem. The specification with the best AIC has a log-likelihood of $-0.37$ per observation, improving on the linear additive model ($LL_{\mathrm{linear}}=-0.46$). Its performance is also comparable to the value of $-0.36$ reported for the MNL specification developed by expert modellers in \cite{tsoleridis2022deriving}, as shown in Table~\ref{tab:decisions_spec_and_fit}. The Delphos specification includes alternative-specific constants, non-linear transformations, and interactions with socio-demographic variables.

Although the two datasets differ substantially in sample size, choice-set structure, and available variables, the same transferred policy identifies competitive specifications for both. Delphos achieves a higher log-likelihood per observation than the VNS benchmark on Swissmetro and performance comparable to a published MNL specification developed by expert modellers on Decisions.

\begin{table}[ht]
\centering
\caption{Best specification identified by Delphos on Decisions.}
\label{tab:decisions_spec_and_fit}
\small
\begin{tabular}{lcccc}
\hline
Modelling term & Included & Taste & Transformation & Interaction \\
\hline
ASC         & \checkmark & Specific & --       & Purpose  \\
Travel time & \checkmark & Specific & Linear & N. Car \\
Travel cost & \checkmark & Generic & Box-Cox   & N. Car\\
Headway     & \checkmark & Specific  & Log      & Gender \\
\hline
LL($0$) & \multicolumn{4}{r}{-11,233.44} \\
LL(final) & \multicolumn{4}{r}{-3,481.79} \\
AIC & \multicolumn{4}{r}{7,055.59} \\
BIC & \multicolumn{4}{r}{7,384.20} \\
Rho-squared & \multicolumn{4}{r}{0.69} \\
Adj.\ Rho-squared & \multicolumn{4}{r}{0.69} \\
Observations & \multicolumn{4}{r}{9,356} \\
Number of parameters & \multicolumn{4}{r}{46} \\\hline
$LL/N$ -- Delphos & \multicolumn{4}{r}{-0.37} \\
$LL/N$ -- \cite{tsoleridis2022deriving} & \multicolumn{4}{r}{-0.36} \\
\hline
\end{tabular}
\end{table}

\section{Conclusions}\label{sec:conclusions}
This paper extends Delphos to a multitask reinforcement learning setting for assisted discrete choice model specification across related transport datasets. We formulate specification as a sequential decision problem in which an agent builds a candidate through modelling actions and submits it to an estimation environment. A DeepSet-Q architecture represents each candidate as a set of modelling terms and combines this representation with dataset-specific context. This design allows the same decision rule to be applied across datasets with different variables.\\

Our results show that Delphos learns reusable modelling strategies that transfer across specification problems. During training, the multitask agent consistently outperforms independently trained single-task agents. It achieves higher rewards, greater sample efficiency, and a larger proportion of successfully estimated specifications while exploring fewer candidates. These results suggest that shared modelling experience guides the search towards promising specification trajectories rather than simply encouraging broader exploration. When applied without further training to the unseen Swissmetro and Decisions datasets, the same agent identifies competitive and behaviourally meaningful specifications in less than 20 minutes on a standard CPU. It achieves a higher log-likelihood per observation than the VNS benchmark on Swissmetro and performance comparable to a published MNL specification developed by expert modellers on Decisions. Overall, these findings highlight the potential of reinforcement learning to accumulate and reuse experience for assisted choice model specification.\\

Beyond these results, the framework can reduce the cognitive and computational burden of manual trial and error and provide suggestions for behaviourally sound utility specifications. This support may help reduce the risk of misspecification, which can bias parameter estimates, weaken forecasts, and lead to misleading welfare conclusions. Delphos may therefore help choice modellers build utility functions that can be used to derive marginal effects and willingness-to-pay measures, including values of travel time savings for cost-benefit analysis.\\

The multitask training setup has several limitations. First, Delphos is trained on only nine datasets and evaluated on two unseen datasets. Although these datasets have distinct catalogues, broader validation across additional problems is needed to assess the generality of the learnt policy. Second, the reward function is based mainly on goodness-of-fit and convergence. It could be extended to include explicit behavioural constraints, as in \citep{nova2025delphos}. Finally, the current framework transfers a fixed policy without adapting it to new datasets during inference.\\

Delphos can be then used as a software package for assisted choice model specification and estimation. The modeller first defines a catalogue of attributes, socio-demographic characteristics, non-linear transformations, and taste structures. The agent then proposes and returns successfully estimated utility specifications on the Pareto front. The modeller can compare these models and select promising candidates for further diagnosis, refinement, and validation. This workflow automates part of the search while retaining modeller control over interpretation and final selection.\\

Future work will extend the experimental setup to a larger set of datasets to assess transferability across a broader range of modelling contexts. We will also investigate context representations that describe each problem using dataset statistics and behavioural features, rather than only the availability of catalogue components. These representations may enable more informed decisions across heterogeneous datasets. Finally, we will evaluate few-shot adaptation strategies to determine when a shared policy can be efficiently specialised to a new dataset using only a small number of additional training episodes.\\

\section*{Acknowledgements}
Stephane Hess acknowledges support from the European Research Council through the Advanced Grant 101020940-SYNERGY.

\clearpage
\section*{Appendix}
\begin{table}[ht]
\centering
\caption{Hyperparameter configuration used for DeepSet-Q.}
\label{tab:hyperparameters}
\small
\begin{tabular}{ll}
\toprule
\textbf{Domain catalogue} & \\
\midrule
Attributes $(|\mathcal{K}|)$           & 7 \\
Transformations $(|\mathcal{T}|)$      & 3 \\
Taste structures $(|\mathcal{G}|)$     & 2 \\
Covariates $(|\mathcal{V}|)$           & 7 \\
Training datasets                      & 9 \\
\midrule
\textbf{DeepSet encoder} & \\
\midrule
Input modelling term              & $(k,t,g,v)$ \\
Embedding dimensions              & $(16,8,8,16)$ \\
Hidden layer 1                    & 128 \\
Hidden layer 2                    & 64 \\
Modelling term ($d_{\text{term}}$) & 64 \\
Pooling                           & Mean \\
Specification embedding ($Z(s)$)  & 64 \\
Activation                        & ReLU \\
Weight initialisation             & Kaiming uniform \\
Trainable parameters              & 27,384 \\
\midrule
\textbf{Deep Q-network} & \\
\midrule
Specification embedding          & 64 \\
Context vector                   & 19 \\
Input dimension                  & 83 \\
Hidden layer 1                   & 256 \\
Hidden layer 2                   & 256 \\
Output actions                   & 297 \\
Activation                       & ReLU \\
Weight initialisation            & Kaiming uniform \\
Trainable parameters             & 158,761 \\
\midrule
\textbf{Training} & \textbf{Value} \\
\midrule
Episodes per task                 & 10,000 \\
Learning updates                  & 10,000 \\
Replay buffer sampling            & 10,000 \\
Mini-batch size                   & 64 \\
Discount factor ($\gamma$)        & 0.90 \\
Learning rate                     & $1\times10^{-3}$ \\
Target update ($\tau$)            & 0.005 \\
$\epsilon$-greedy exploration     & Linear decay \\
\bottomrule
\end{tabular}
\end{table}

\clearpage


\end{document}